\documentclass[a4paper,12pt]{article}
\usepackage{amsmath,amsfonts,amssymb,amsthm,amstext,amscd}
\usepackage{latexsym}
\usepackage{hyperref}
\usepackage{caption}
\usepackage{comment}
\usepackage{graphicx}
\usepackage{color}
\usepackage{setspace}
\usepackage{xcolor}
\usepackage{titletoc}
\usepackage{cancel}

\usepackage[normalem]{ulem}
\usepackage{cite}

\newcommand{\mrd}{\mathrm{d}}

\renewcommand*{\eqref}[1]{%
	\begingroup
	\hypersetup{
		linkcolor=black,
		linkbordercolor=black,
	}%
	\textcolor{black}{(\ref{#1})}%
	\endgroup
}
\hypersetup{linkcolor=black,citecolor=black,urlcolor=black,colorlinks=true}

\begin{document}

\begin{titlepage}
\vspace{0.5cm}
\begin{center}
\renewcommand{\baselinestretch}{1.5}
\setstretch{1.7}
{\fontsize{20pt}{20pt}\selectfont An Exact Single-Rotating Near-Horizon Geometry in Einstein-Gauss-Bonnet Gravity}

\vspace{9mm}
\renewcommand{\baselinestretch}{1}  
\setstretch{1}

\centerline{\large{ {U. Can Çelik,$^{\ast}$\footnote{ucelik@metu.edu.tr}} {Kamal Hajian,$^{\ast\dagger}$\footnote{kamal.hajian@uni-oldenburg.de}} and {Jutta Kunz$^{\dagger}$\footnote{jutta.kunz@uni-oldenburg.de}} }}
\vspace{4mm}
\normalsize
$^\ast$\textit{Department of Physics, Middle East Technical University, 06800, Ankara, Turkey}\\
$^\dagger$\textit{Institute of Physics, University of Oldenburg, P.O. Box 2503, D-26111 Oldenburg, Germany}\\
\vspace{5mm}

\begin{abstract}\noindent
We construct a five-dimensional singly rotating near-horizon solution in Einstein-Gauss-Bonnet gravity. We show that the Gauss-Bonnet term removes the local curvature singularity, yielding finite curvature invariants throughout the spacetime, provided the rotation parameter remains below a certain value set by the Gauss-Bonnet coupling. To our knowledge, this is the first analytic example of a singly rotating five-dimensional solution in this framework with finite curvature invariants over a nontrivial region of parameter space. We analyze the geometry across this space, identifying regular noncompact, singular, and marginal regimes. Finally, we study the thermodynamic properties, finding that while higher-derivative corrections regularize the local curvature behavior, they also introduce unique challenges to the standard thermodynamic description of Killing horizons.
\end{abstract}

\end{center}

\let\newpage\relax

\end{titlepage}

\section{Introduction}
Lovelock gravity provides the most general extension of Einstein gravity in higher dimensions that yields second-order field equations for the metric and avoids the occurrence of Ostrogradsky instabilities \cite{Lovelock:1971yv}.
Originally formulated by Lovelock, it consists of a series of higher-curvature terms, each one contributing non-trivially only above a certain spacetime dimension \cite{Lovelock:1971yv,Lovelock:1972vz}.
Among these, the Gauss-Bonnet (GB) term represents the leading correction beyond the Einstein-Hilbert action of general relativity, and is non-trivial already in five dimensions.
Moreover, it plays a distinguished role, as it arises naturally in the low-energy effective action of string theory \cite{Zwiebach:1985uq}.

Black hole solutions in Einstein-Gauss-Bonnet (EGB) gravity were first obtained by Boulware and Deser \cite{Boulware:1985wk}.
These static solutions are characterized by the presence of two distinct branches that arise from the quadratic structure of the field equations. 
The physically relevant (Einstein) branch admits a single event horizon for all positive masses.
In addition, the theory introduces a novel branch singularity associated with the higher-curvature corrections.
The physically relevant Einstein branch avoids this singularity, while it is associated with the non-Einstein (Gauss-Bonnet) branch and signals a breakdown of the solution at finite radius \cite{Boulware:1985wk,Cai:2001dz}.

Rotating black holes in higher dimensions were originally discovered by Myers and Perry \cite{Myers:1986un}.
However, the inclusion of rotation in EGB gravity remains technically challenging, and exact solutions are generally not known.
Most existing results rely on numerical methods, through which black holes with equal angular momenta as well as singly rotating black holes have been constructed \cite{Brihaye:2010wx,Kleihaus:2023clx}.
These results show that GB corrections significantly alter the domain of existence; in particular, at fixed mass the solutions studied in  \cite{Kleihaus:2023clx} exist only up to a maximal value of the coupling constant, while for spherical horizons the Einstein-gravity angular-momentum bound is not exceeded.

While most studies of rotating EGB black holes focus on global solutions, the near-horizon limit of extremal geometries provides a complementary framework in which the effects of higher-curvature corrections can often be analyzed more explicitly.
Near-horizon extremal geometries (NHEGs) provide a powerful framework for studying extremal black holes, as they exhibit an enhanced $\mathrm{SL}(2,\mathbb{R}) \times \mathrm{U}(1)^n$ isometry group (where $n$ represents the number of axial isometry directions) \cite{Bardeen:1999px,Kunduri:2013gce,Kunduri:2007vf,Kunduri:2008rs}.
These geometries capture universal features of extremal horizons.
Recently, a single-rotating near-horizon solution in five-dimensional Einstein gravity was constructed, which, however, exhibits a curvature singularity at the poles of the angular coordinates \cite{Hajian:2023gsc}.
The presence of GB corrections modifies the near-horizon geometry through additional curvature contributions to the field equations.
In particular, since higher-curvature terms become significant in regions of large curvature, they can affect the regularity properties of near-horizon geometries \cite{Sen:2005wa,Callan:1988hs}.

This motivates the study of near-horizon geometries in EGB gravity as a testing ground for the modification of singular structures by higher-curvature corrections. Here, we construct an exact analytic near-horizon solution describing a five-dimensional single-rotating extremal geometry in EGB gravity.
We show that the inclusion of the GB term removes the local curvature singularity present in the corresponding Einstein gravity solution \cite{Hajian:2023gsc}, rendering curvature invariants finite throughout the spacetime within a finite parameter domain. 
However, the resulting geometry also exhibits unconventional global features, including divergences in the horizon area and associated conserved charges, indicating that the thermodynamic interpretation of the solution requires further investigation. This suggests that higher-derivative corrections can soften or remove certain local curvature singularities present in the general relativity limit.

It should be emphasized that our analysis is restricted to the near-horizon geometry, and the existence of a corresponding global black hole solution remains an open question. 
Nevertheless, our results provide further evidence that higher-curvature corrections 
may play an important role in modifying the singular structure of extremal rotating geometries and may offer insight into the resolution of singularities in more general gravitational settings. To our knowledge, this represents the first exact analytic singly rotating near-horizon solution in five-dimensional EGB gravity with finite local curvature invariants within a nontrivial parameter domain.

The remainder of this paper is organized as follows. 
In Sec.~2, we introduce the EGB framework and present the exact near-horizon solution for the singly rotating case
and discuss its symmetry structure. 
In Sec.~3, we analyze its geometric properties. 
In Sec.~4, we investigate the associated thermodynamic quantities and discuss the emergence of divergent horizon charges.
Finally, in Sec.~5, we summarize our results and outline possible directions for future work.

\section{The solution}

In five-dimensional gravity, higher-curvature corrections such as the GB term can significantly modify the structure of extremal near-horizon geometries
\cite{Brihaye:2010wx}. 
Motivated by the possibility that higher-derivative corrections may soften the singular behavior of single-rotating solutions \cite{Brihaye:2010wx}, we consider EGB theory. 
The Lagrangian density is given by:
\begin{align}\label{Lagrangian}
\mathcal{L} = \frac{1}{16\pi G}\left( R +\alpha \mathcal{L}_{\text{GB}}\right), \qquad  \mathcal{L}_{\text{GB}}= R^2 - 4R_{\mu\nu}R^{\mu\nu} + R_{\mu\nu\sigma\tau}R^{\mu\nu\sigma\tau},
\end{align}
where $R^\mu_{\, \nu\sigma\tau}$, $R_{\mu\nu}$ and $R$ denote the Riemann tensor, Ricci tensor, and curvature scalar, respectively. The corresponding field equation is:
\begin{equation}
R_{\mu\nu} - \frac{1}{2} R g_{\mu\nu} + \alpha H_{\mu\nu} = 0,
\end{equation}
where
\begin{equation}
H_{\mu\nu} = 2(R_{\mu\sigma\kappa\tau}{R_{\nu}}^{\sigma\kappa\tau} - 2R_{\mu\rho\nu\sigma}R^{\rho\sigma} - 2R_{\mu\sigma}{R^{\sigma}}_{\nu} + R R_{\mu\nu}) - \frac{1}{2}\mathcal{L}_{\text{GB}}g_{\mu\nu}.
\end{equation}

NHEGs generally exhibit an enhanced isometry group \cite{Bardeen:1999px,Kunduri:2013gce}. In five-dimensional spacetime, it is $\text{SL}(2,\mathbb{R}) \times \text{U}(1)^2$. For such a spacetime with a single non-vanishing angular momentum, we propose the following analytic solution:
\begin{equation}
\mrd s^2= \frac{\Delta_+}{\cos^2\theta}\left(\frac{\mrd r^2 }{4r^2}+  \left(\frac{\Delta_-}{\Delta_+}\right)^2\mrd \theta^2+\frac{a^2\sin^2\theta}{\Delta_+}\mrd \phi^2+\mrd \psi^2+ r\mrd t \mrd \psi\right),
\end{equation}
where
\begin{equation}
\Delta_+=a^2\cos^4\theta+2\alpha, \qquad \Delta_-=a^2\cos^4\theta-2\alpha.     
\end{equation}
As a convention, we have chosen the non-vanishing angular momentum to be aligned with the $\psi$-direction.

It is customary to express such solutions in the standard NHEG metric ansatz using the coordinates $(t,r,\theta,\varphi^1,\varphi^2)$, with the identifications $\varphi^1=\phi$ and $\varphi^2=\psi$, as follows:
\begin{align}
{\mrd s}^2=\Gamma(\theta)\left(-r^2 \mrd t^2+\frac{\mrd r^2}{r^2}+4\sigma(\theta)\mrd\theta^2+\sum_{i,j=1}^2\gamma_{ij}(\theta)(\mrd\varphi^i+k^ir\mrd t)(\mrd\varphi^j+k^jr\mrd t)\right),\label{NHEG} \nonumber\\
\Gamma= \frac{\Delta_+}{4\cos^2\theta}, \qquad \sigma =\left(\frac{\Delta_-}{\Delta_+}\right)^2, \qquad \gamma_{11}=\frac{4a^2\sin^2\theta}{\Delta_+}, \qquad \gamma_{22}=4, \quad \gamma_{12}=0
\end{align}
with $k^1=0$ and $k^2=\frac{1}{2}$. The first two terms in the metric \eqref{NHEG} describe an AdS$_2$ space in the Poincar\'{e} patch with $r=0$ as the Poincar\'{e} horizon. In these coordinates, the $\text{SL}(2,\mathbb{R})\times\text{U}(1)^2$ isometry generators are
\begin{equation}
\xi_- =\partial_t\,,\qquad \xi_0=t\partial_t-r\partial_r,\qquad	\xi_+ =\dfrac{1}{2}(t^2+\frac{1}{r^2})\partial_t-tr\partial_r-\frac{1}{r}{k}^i{\partial}_{\varphi^i}, \quad \mathrm{m}_i=\partial_{\varphi^i}
\end{equation}
with the commutation relations 
\begin{align}\label{commutation relation}
[\xi_0,\xi_-]=-\xi_-,\qquad [\xi_0,\xi_+]=\xi_+, \qquad [\xi_-,\xi_+]=\xi_0\,,\qquad [\xi_\mathrm{a},\mathrm{m}_i]=0, \quad \mathrm{a}\in \{-,0,+\}.
\end{align}
The $(t,r,\varphi^2)$ sector of the metric is locally a self-dual orbifold of $\mathrm{AdS}_3$, fibered over the polar direction $\theta$. After putting the metric in NHEG form, this part appears as
\begin{equation}
-r^2 \mrd t^2+\frac{\mrd r^2}{r^2}+4\left(\mrd\varphi^2+k^2\, r\, \mrd t\right)^2,
\end{equation}
which is the standard representation of $\mathrm{AdS}_3$ written as an $S^1$ fiber over $\mathrm{AdS}_2$. The constant $k^2$ fixes the quotient: for $k^2\neq 0$, the $\varphi^2$-circle is non-trivially fibered over the AdS$_2$ base, producing a discrete identification along a null direction of $\mathrm{AdS}_3$. This is precisely what defines the self-dual orbifold—one quotients global $\mathrm{AdS}_3$ by a lightlike combination of its $\text{SL}(2,\mathbb{R})\times \text{SL}(2,\mathbb{R})$ isometries, leaving a geometry that still locally has constant negative curvature but globally has a compact null direction. In the present solution, this structure is warped by the function $\Gamma(\theta)$, so the size of the fiber and the effective AdS$_2$ scale vary over the transverse space, but the local $\mathrm{AdS}_3$ (and hence $\text{SL}(2,\mathbb{R})\times \text{U}(1)$) symmetry of the $(t,r,\varphi^2)$ sector is preserved. This is the hallmark of near-horizon extremal geometries: the spacetime organizes itself as a fibration of a (possibly orbifolded) $\mathrm{AdS}_3$  over the transverse space, with the self-dual orbifold capturing the chiral, or ``single Virasoro'' structure that often underlies their holographic description \cite{Kerr/CFT}.

\section{Some geometric features}
In this section, we explore the fundamental geometric properties of the new geometry constructed in Sec.~2. Our analysis focuses on characterizing the local curvature and the global regularity requirements of the spacetime. To evaluate the regularity of the geometry and identify potential physical singularities, we first compute the Kretschmann scalar, $K \equiv R_{\mu\nu\alpha\beta}R^{\mu\nu\alpha\beta}$, which is found to be 
\begin{equation}\label{Kretschmann}
    K = 4\left(\frac{2\Delta_{+}^{2}+2\Delta_{+}\Delta_{-}-\Delta_{-}^{2}}{\Delta_{-}^{3}}\right)^{2}+\frac{24}{\Delta_{-}^{2}}+\frac{12}{\Delta_{+}^{2}}.
\end{equation}
The behavior of this invariant is governed by the metric functions $\Delta_+$ and $\Delta_-$. As we shall see in the subsequent subsections, the presence or absence of roots for these functions within the coordinate range determines the domain of existence for regular solutions.

In addition to local curvature regularity, the global consistency of the spacetime requires the absence of conical singularities. In the present geometry, such a singularity may arise at the pole $\theta=0$. To ensure a smooth geometry at this location, the periodicity of the angular coordinate $\phi$ must be fixed to
\begin{equation}\label{Del varphi}
    \Delta\phi=2\pi\frac{|a^{2}-2\alpha|}{\sqrt{a^{2}(a^{2}+2\alpha)}}.
\end{equation}
This condition illustrates how the GB coupling $\alpha$ and the rotation parameter $a$ non-trivially modify the geometric properties of the horizon compared to the pure Einstein gravity case. 
With these basic geometric quantities established, we now turn to a detailed analysis of the parameter space, distinguishing between regular, singular, and marginal domains.

\subsection{Regular noncompact domain $a^2<2|\alpha|$}
A notable property of this EGB metric is that its local curvature invariants remain finite provided
\begin{equation}
a^2<2|\alpha|.
\end{equation}
This is the region where neither $\Delta_+$ nor $\Delta_-$ has any roots. In this domain of parameters, evaluating the Kretschmann scalar at the pole $\theta=\frac{\pi}{2}$ yields
\begin{equation}
\lim_{\theta\to\frac{\pi}{2}}K=\frac{10}{\alpha^2},
\end{equation}
which is finite in the limit $\theta\to\frac{\pi}{2}$ for any non-zero GB coupling $\alpha$. 
This demonstrates that the higher-derivative corrections remove the local curvature singularity previously observed in the extremal limit of Myers-Perry generalizations 
\cite{Brihaye:2010wx,Hajian:2023gsc}. Explicitly, in the pure Einstein gravity limit ($\alpha\to 0$ while $a\neq 0$), the metric reduces to the Einstein single-rotating solution presented in Ref. %
\cite{Hajian:2023gsc}:
\begin{equation}
\mrd s^2= a^2\cos^2\theta\left(\frac{\mrd r^2 }{4r^2}+  \mrd \theta^2+\frac{\sin^2\theta}{\cos^4\theta}\mrd \phi^2+\mrd \psi^2+ r\mrd t \mrd \psi\right),
\end{equation}
while the Kretschmann scalar reduces to 
\begin{equation}
\lim_{\alpha\to 0}K=\frac{72}{a^4\cos^8\theta},
\end{equation}
indicating a curvature singularity at $\theta\to \frac{\pi}{2}$ %
\cite{Hajian:2023gsc}. To set the stage, it is useful to review the pure Einstein gravity analysis of a single rotating extremal Myers-Perry black hole in five dimensions, which features a singular horizon with a vanishing area. Two distinct approaches exist for studying the near-horizon geometry of this black hole. The first, known as the extremal vanishing horizon (EVH) limit \cite{Fareghbal:2008ar,Fareghbal:2008eh,Sheikh-Jabbari:2011sar,Sadeghian:2014tsa,Sadeghian:2015laa,Sadeghian:2015nfi,Noorbakhsh:2017nde}, keeps the entropy at zero while simultaneously setting the sole non-zero angular momentum to zero. The second approach preserves the angular momentum while yielding a non-zero entropy. The solution presented in this paper connects to the latter approach in the limit as $\alpha \to 0$.  

While the finiteness of the Kretschmann scalar at $\theta=\pi/2$ indicates the absence of a local curvature singularity, it is crucial to emphasize that finite curvature invariants are not sufficient to guarantee a fully regular spacetime. Specifically, the norm of the Killing vector field $\partial_{\psi}$, given by the metric component $g_{\psi\psi} = \frac{\Delta_{+}}{\cos^{2}\theta}$, diverges as $\theta \rightarrow \pi/2$. At this limit, the numerator $\Delta_{+}$ approaches a finite value of $2\alpha$, while the denominator vanishes, causing the spacetime structure to break down if one insists on a compact spatial section. Consequently, this geometry cannot be interpreted as a near-horizon limit with a standard compact 3-sphere topology. To ensure a strictly non-divergent metric, the physical domain of the polar coordinate must be restricted to the half-open interval $\theta \in [0, \pi/2)$. Under this restriction, the asymptotic region $\theta \rightarrow \pi/2$ effectively acts as a conformal boundary, implying that the near-horizon geometry inherently possesses non-compact spatial sections.

In contrast to the usual situation for angular coordinates, the periodicity of $\psi$ cannot generically be fixed by a conical-regularity argument, since this requires the existence of an axis where the Killing field $\partial_\psi$ vanishes. In the present geometry, the norm of this Killing vector is $g_{\psi\psi}$. Whether an axis exists depends on the parameters. 
For $a^2<2|\alpha|$, one has $\Delta_+\neq 0$ for all $\theta$, and hence $g_{\psi\psi}$ never vanishes, so the $\psi$-circle does not shrink anywhere. In this case, there is therefore no local geometric condition that fixes the periodicity of $\psi$, which must instead be specified by global considerations, such as matching to an underlying spacetime or choosing a particular $\text{U}(1)$ bundle structure.

It is important to note the implications of a negative GB coupling. Although the condition $a^2<2|\alpha|$ guarantees that neither $\Delta_+$ nor $\Delta_-$ has a zero in the interval $\theta\in[0,\pi/2]$, for $\alpha<0$ one has
\begin{equation}
\Delta_+ = a^2\cos^4\theta - 2|\alpha| <0
\end{equation}
throughout this domain. Consequently, the overall conformal factor $\Gamma$ becomes negative, and the metric no longer has the same Lorentzian signature as in the $\alpha>0$ branch. Moreover, the regularity condition for the $\phi$-circle in \eqref{Del varphi} becomes ill-defined in this range, since $a^2+2\alpha=a^2-2|\alpha|<0$. Therefore, although the local curvature invariant $K$ remains finite, the negative
coupling branch does not define a regular physical near-horizon geometry with the desired signature and angular periodicity. So, the regular domain should therefore be understood as the positive-coupling branch
\begin{equation}\label{regular}
\alpha>0,\qquad a^2<2\alpha .
\end{equation}
Note that this domain does not connect smoothly to the Einstein branch: taking the limit $\alpha\to 0$ under the above condition forces $a\to0$, resulting in a completely singular solution that lacks any free physical parameters.

\subsection{Singular domain $a^2>2|\alpha|$}
When $a^2 > 2|\alpha|$, a curvature singularity occurs at the zero of either $\Delta_+$ or $\Delta_-$, depending on the sign of $\alpha$: specifically, if $\alpha > 0$, only $\Delta_-$ can vanish, and conversely. The singularity follows directly from the fact that each of the three terms in \eqref{Kretschmann} is positive. Consequently, if any one of these terms diverges, the Kretschmann scalar $K$ also diverges.

In this singular domain, if $\alpha<0$, the function $\Delta_+$ can vanish at some $\theta$, implying $g_{\psi\psi}=0$ and the existence of a would-be axis of the $\psi$-circle. In that case, one may attempt to fix the periodicity of $\psi$ by demanding the absence of a conical singularity. The period of $\psi$ is found to be
\begin{equation}
\Delta\psi=2\pi\frac{\left(\frac{2|\alpha|}{a^2}\right)^{1/4}}{\sqrt{1-\left(\frac{2|\alpha|}{a^2}\right)^{1/2}}}. 
\end{equation}

\subsection{Marginal domain $a^2=2|\alpha|$}
To fully understand the physical nature of this boundary, we must analyze the roots of the metric functions $\Delta_{+}$ and $\Delta_{-}$ at exactly $a^{2}=2|\alpha|$. The behavior depends strictly on the sign of the GB coupling $\alpha$:
\begin{itemize}
    \item For $\alpha > 0$ (yielding $a^{2}=2\alpha$): The function $\Delta_{-}$ becomes $a^2(\cos^{4}\theta - 1)$. At the location $\theta = 0$, $\Delta_{-}$ vanishes.
    \item For $\alpha < 0$ (yielding $a^{2}=-2\alpha$): The function $\Delta_{+}$ becomes $a^2(\cos^{4}\theta - 1)$. At the location $\theta = 0$, $\Delta_{+}$ vanishes.
\end{itemize}
While the vanishing of the $\mrd\theta^{2}$ component signals a breakdown in the metric signature, examining the Kretschmann scalar proves that this boundary represents a much more severe physical pathology. Because the terms in the Kretschmann scalar are strictly positive and feature both $\Delta_{-}$ and $\Delta_{+}$ in their denominators, the scalar diverges if either function goes to zero. Consequently, as $\theta \rightarrow 0$ in the marginal domain, $K \rightarrow \infty$.

Therefore, the boundary case $a^{2}=2|\alpha|$ does not merely represent a coordinate singularity; it manifests as a true, unavoidable physical curvature singularity at $\theta = 0$. This confirms that the regular, singularity-free spacetime is strictly confined to the open parameter space in \eqref{regular}, and the introduction of GB corrections can only regularize the near-horizon extremal geometry within this specific bound. 

This physical breakdown at the marginal boundary is also directly reflected in the behavior of the periodicity $\Delta\phi$ derived in Eq. \eqref{Del varphi}. In the limit where $a^{2} \rightarrow 2\alpha$ (for $\alpha > 0$), the numerator of the periodicity condition vanishes, leading to $\Delta\phi \to 0$, which implies the collapse of the $\phi$-circle. Conversely, in the case where $a^{2} \rightarrow -2\alpha$ (for $\alpha < 0$), the denominator of the expression vanishes, causing the periodicity to diverge. These pathological behaviors of the angular coordinate's period at the boundary $a^{2}=2|\alpha|$ serve as a global manifestation of the local curvature singularity occurring at the pole $\theta=0$, further confirming that the marginal case represents the limit of physical validity for the solution.

\section{Some thermodynamic features}
To study the thermodynamics of the NHEGs in $D$ dimensions it is helpful to define the surfaces $\mathcal{H}$ as $(D-2)$-dimensional closed, smooth manifolds identified by constant time and radius coordinates $(t_{\mathcal{H}}, r_{\mathcal{H}})$, serving as the analogue of the bifurcation surface in black holes \cite{Hajian:2013lna,Hajian:2014twa}. 
Because these surfaces are mapped to one another through the action of the $\{\xi_{-}, \xi_{0}\}$ isometry subalgebra, they are related by the spacetime symmetries, which guarantees that thermodynamic variables such as the entropy $S$ and conserved angular momenta $J_{i}$ remain independent of the particular choice of integration surface $\mathcal{H}$. Within the NHEG phase space, these surfaces are mathematically constructed to preserve their smoothness and constant area, ensuring that the geometric properties of the ``throat" remain stable across various metric configurations. In Appendix, a review of the NHEG conserved charges and thermodynamics is provided. 

In the context of singular NHEG solutions, it is important to note that the systematic analysis of the charges  relies on an implicit assumption: the smoothness of the integration surfaces $\mathcal{H}$. Our results indicate that the standard NHEG thermodynamics is not applicable to the solution presented in this work, even in the regular regime $a^2 < 2|\alpha|$ with $0<\alpha$ and $\theta \in [0, \pi/2)$.  To be more specific, consider the area of $\mathcal{H}$, obtained from the integration over the square root of determinant of the $(\theta,\phi,\psi)$ block of the metric:
\begin{align}
A_\mathcal{H}&=\oint_\mathcal{H}  \sqrt{g_{\theta\theta}g_{\phi\phi}g_{\psi\psi}}=\int\mrd \psi\, \mrd \phi \, \mrd \theta \, \frac{a \sin \theta (2\alpha - a^2 \cos^4 \theta)}{\cos^3 \theta} \\
&=\int_{0}^{\Delta\psi}\mrd \psi\int_{0}^{\Delta\phi}\mrd \phi  \left(\frac{a^3\cos^2\theta}{2} +\frac{a\alpha}{\cos^2\theta} \right)\Big|_{\theta=0}^{\theta=\frac{\pi}{2}}.
\end{align}
We observe that the last term in the integrand diverges as $\theta \to \frac{\pi}{2}$, rendering the area infinite. Given the divergence of the area element $A_{\mathcal{H}}$ at the conformal end $\theta \rightarrow \frac{\pi}{2}$, the standard definitions of entropy $S$ and the second angular momentum $J_{2}$ result in ill-defined, infinite values even within the ostensibly regular parameter regime $a^{2} < 2|\alpha|$. This behavior suggests that the standard thermodynamic framework for NHEGs requires a more refined analysis when higher-curvature corrections are present. Specifically, in the regular sector, the charge densities should be studied because the associated integral surfaces $\mathcal{H}$ are non-compact. 

In the standard NHEG thermodynamics, the generator of the entropy is provided by the Killing vector field $\zeta_{\mathcal{H}}$, which vanishes on the bifurcation surface and becomes null on the corresponding Killing horizon. This vector is expressed as a linear combination of the SL$(2,\mathbb{R})\times \text{U}(1)^{2}$ generators \cite{Hajian:2013lna}:
\begin{equation}\label{zetaH}
    \zeta_{\mathcal{H}} = n_{\mathcal{H}}^{\mathrm{a}} \xi_\mathrm{a} - k^{i} \mathrm{m}_i,
\end{equation}
where $n_{\mathcal{H}}^{\mathrm{a}}$ are constants determined by the choice of the bifurcation point $(t_{\mathcal{H}}, r_{\mathcal{H}})$ (see Ref. \cite{Hajian:2015eha} for their physical interpretation) 
\begin{align}
    n_{\mathcal{H}}^{-} = -\frac{t_{\mathcal{H}}^{2}r_{\mathcal{H}}^{2}-1}{2r_{\mathcal{H}}}, \qquad
    n_{\mathcal{H}}^{0} = t_{\mathcal{H}}r_{\mathcal{H}}, \qquad 
    n_{\mathcal{H}}^{+} = -r_{\mathcal{H}}.
\end{align}
The surface gravity associated with these horizons in the standard NHEG geometry is normalized to $\kappa=1$.

By the linearity of the covariant phase space charge variations with respect to the Killing generators $ 2\pi \zeta_{\mathcal{H}}$, $\xi_\mathrm{a}$, and $-\mathrm{m}_i$ (which are the generators of the entropy density $\mathcal{S}$, SL$(2,\mathbb{R})$ charges $\mathcal{Q}_\mathrm{a}$, and angular momenta $\mathcal{J}_i$ respectively),  the same fundamental relation as Eq. \eqref{zetaH} between the charge density variations would follow,
\begin{equation}\label{New pert law}
\frac{\delta \mathcal{S}}{2\pi} =n_{\mathcal{H}}^{\mathrm{a}} \delta \mathcal{Q}_\mathrm{a}+ k^i \delta \mathcal{J}_i.
\end{equation}
Consequently, one obtains the density counterpart of the standard NHEG entropy perturbation law (see Eq.\eqref{NHEG entropy perturbation}), but with non-vanishing contribution from the $\mathrm{SL}(2,\mathbb{R})$ charge densities. In the standard NHEG dynamics, this latter contribution vanishes via the integration over smooth and compact $\mathcal{H}$ surface. 

 The relation \eqref{New pert law} can be checked explicitly via the variation of the metric with respect to the parameter $a$, which yields
\begin{align}
\frac{\delta \mathcal{S}}{2\pi}&=\delta\mathcal{J}_2=2n_{\mathcal{H}}^{\mathrm{a}} \delta\mathcal{Q}_\mathrm{a}=\frac{\sin(\theta)\mathcal{P}(\Delta_+,\Delta_-)}
{512\pi G\cos^3(\theta)\Delta_-^3\Delta_+^2}\delta a
\end{align}
with
{\footnotesize
\begin{align}
\mathcal{P}=
419\Delta_-^6 +10730\Delta_-^5\Delta_+ -40043\Delta_-^4\Delta_+^2 +46828\Delta_-^3\Delta_+^3
-14627\Delta_-^2\Delta_+^4 -6262 \Delta_{-} \Delta_+^5 +3147\Delta_+^6 .
\end{align}}
\noindent Using $k^1=0$, $\mathcal{J}_1=0$, and $k^2=\frac{1}{2}$ one verifies that the relation \eqref{New pert law} is satisfied. Noting that $k^i$ and $n_{\mathcal{H}}^{\mathrm{a}}$ are constants independent of the parameter $a$, we can integrate the perturbation law \eqref{New pert law} to find the finite charge-density relation 
\begin{equation}\label{New entropy law}
\frac{\mathcal{S}}{2\pi} =n_{\mathcal{H}}^{\mathrm{a}} \mathcal{Q}_\mathrm{a}+ k^i \mathcal{J}_i
\end{equation}
up to an integration constant which is independent of the parameter $a$.

\section{Conclusions}
In this work, we have presented an exact analytic near-horizon extremal geometry (NHEG) solution in five-dimensional Einstein-Gauss-Bonnet gravity. The construction  may be viewed as a higher-curvature generalization of the single-rotating NHEG solution in general relativity, which is otherwise characterized by a curvature singularity at the poles. 
We have demonstrated that the inclusion of the Gauss-Bonnet term removes the local curvature singularity present in the Einstein gravity solution, rendering local curvature invariants finite throughout the geometry within a finite parameter domain, provided that the rotation parameter is below a certain value which is determined by the Gauss-Bonnet coupling.
Within this regular noncompact domain, the Kretschmann scalar remains finite everywhere, including the limit where the polar coordinate approaches the boundary previously found to be singular in pure Einstein gravity. 
By contrast, in the singular and marginal domains, the metric functions exhibit roots that lead to unavoidable physical curvature singularities.

We have also examined the thermodynamic features of the solution and found that the area element of the horizon diverges at the noncompact conformal end. 
This divergence in turn renders conserved quantities such as the entropy and angular momentum ill-defined within the standard NHEG framework. In the special case $\alpha = 0$, these charges remain finite, the NHEG thermodynamic structure is recovered, and the results reduce to those of general relativity. Because these divergent quantities currently obscure a clear physical interpretation of the solution's phase space, we postpone more detailed thermodynamic studies for later works.

Finally, it should be emphasized that our analysis is strictly restricted to the near-horizon geometry. Whether a corresponding global black hole solution exists that admits this near-horizon geometry as its extremal limit remains an important open question. Future research will be directed toward resolving the thermodynamic divergences identified here and investigating the potential existence of matching global parent spacetimes, taking into account the branch distinction discussed above. Moreover, the coupling $\alpha$ can be treated as a conserved charge for this solution through the introduction of auxiliary scalar and gauge fields \cite{Hajian:2023bhq,Hajian:2025hxf}. Investigating the relationship between the presented solution and the ``vanishing extremal horizon" geometries discussed in the literature (e.g., in Refs. \cite{deBoer:2011zt,Sadeghian:2015laa,Sadeghian:2015nfi}) would be an interesting direction for future work.

\vskip 1 cm

\noindent \textbf{Acknowledgments:} This work has been supported by the TÜBİTAK International Researchers Program No. 2221.

\section*{Appendix}
To evaluate the conserved charges of the solution, we employ the covariant phase space formulation \cite{Wald:1993nt,Iyer:1994ys,IyerWald1995}. This formulation constructs a phase space covariantly over the entire spacetime, bypassing the need for a non-covariant space+time decomposition. In this covariant formalism, the variation of a conserved charge $\delta H_{\xi}$ associated with a symmetry generator (such as a Killing vector $\xi^\nu$) is found by integrating a charge variation $D-2$-form tensor $\boldsymbol{k}_\xi(\delta g_{\mu\nu},g_{\mu\nu})$. Denoting its Hodge dual by $k_{\xi}^{\mu\nu}$, the integration over a surface $\mathcal{H}$ yields
\begin{equation}
\delta H_{\xi}=\oint_{\mathcal{H}}\sqrt{-g} k_{\xi}^{tr}
\end{equation}
in which the integration over the coordinates of $\mathcal{H}$ should be understood. In the EGB theory with the Lagrangian \eqref{Lagrangian}, the total variation tensor decomposes into its pure Einstein gravity and GB contributions (see e.g. Refs. \cite{Ghodrati:2016vvf,Hajian:2015xlp,Hajian:2021hje,Hajian:2025hxf} for more details):
\begin{equation}
k^{\mu\nu} = k_{\text{E}}^{\mu\nu} + \alpha k_{\text{GB}}^{\mu\nu}
\end{equation}
with the Einstein contribution:
\begin{equation}
\begin{aligned}
k_{\text{E}}^{\mu\nu} &= \frac{1}{16\pi G} \Big[ h^{\mu\alpha}\nabla_{\alpha}\xi^{\nu}-\nabla^{\mu}h^{\nu\alpha}\xi_{\alpha}-\frac{1}{2}h\nabla^{\mu}\xi^{\nu}  - (\nabla_{\alpha}h^{\mu\alpha}-\nabla^{\mu}h)\xi^{\nu} \Big] - [\mu\leftrightarrow\nu]
\end{aligned}
\end{equation}
where $h_{\mu\nu} = \delta g_{\mu\nu}$ represents the linearized variations of the metric, and $h \equiv {h^\alpha}_\alpha$.
The GB term further decomposes into three sub-components proportional to the coupling $\alpha$, representing the $R^2$, Ricci-squared, and Riemann-squared terms respectively:
\begin{equation}
k_{\text{GB}}^{\mu\nu} = k_{R^2}^{\mu\nu} - 4k_{\text{Ricci}^2}^{\mu\nu} + k_{\text{Riemann}^2}^{\mu\nu}.
\end{equation}
For completeness, we present the explicit expressions for the higher-curvature contributions to the charge variation tensor:
{\small
\begin{align}
k_{R^2}^{\mu\nu} &= \frac{1}{16\pi G} \Big[ 2R (h^{\mu\alpha}\nabla_{\alpha}\xi^{\nu}-\nabla^{\mu}h^{\nu\alpha}\xi_{\alpha}-\frac{1}{2}h\nabla^{\mu}\xi^{\nu}) \nonumber\\
&\quad + 4(R^{\mu\alpha}\nabla_{\alpha}h-\nabla_{\alpha}Rh^{\mu\alpha}-R_{\alpha}{}^{\mu}\nabla_{\beta}h^{\alpha\beta}-\Box\nabla^{\mu}h+\nabla_{\alpha}\nabla^{\mu}\nabla_{\beta}h^{\alpha\beta} \nonumber\\
&\quad -\nabla^{\mu}(R_{\alpha\beta}h^{\alpha\beta})+\frac{1}{2}\nabla^{\mu}R h)\xi^{\nu} + 2(R_{\alpha\beta}h^{\alpha\beta}-\nabla_{\alpha}\nabla_{\beta}h^{\alpha\beta}+\Box h)\nabla^{\mu}\xi^{\nu} \nonumber\\
&\quad - (2R(\nabla_{\alpha}h^{\mu\alpha}-\nabla^{\mu}h)-2\nabla_{\alpha}R h^{\mu\alpha}+2\nabla^{\mu}R h)\xi^{\nu} \Big] - [\mu\leftrightarrow\nu],
\end{align}
\begin{align}
k_{\text{Ricci}^2}^{\mu\nu} &= \frac{1}{16\pi G} \Big[ (\nabla^{\alpha}R_{\alpha}{}^{\mu}h-\nabla_{\alpha}Rh^{\mu\alpha}-\nabla^{\mu}(R_{\alpha\beta}h^{\alpha\beta})+\nabla^{\mu}\nabla_{\alpha}\nabla_{\beta}h^{\alpha\beta}-\nabla^{\mu}\Box h)\xi^{\nu} \nonumber\\
&\quad + (2\nabla_{\beta}R_{\alpha}{}^{\mu}h^{\beta\nu}-2R^{\mu\beta}\nabla_{\beta}h_{\alpha}{}^{\nu}-2\nabla^{\mu}R_{\alpha\beta}h^{\nu\beta}-\nabla^{\mu}(\nabla_{\alpha}\nabla^{\nu}h-\nabla_{\beta}\nabla_{\alpha}h^{\nu\beta} \nonumber\\
&\quad +\Box h_{\alpha}{}^{\nu}-\nabla^{\beta}\nabla^{\nu}h_{\alpha\beta})+\nabla^{\mu}R_{\alpha}{}^{\nu}h+2R^{\mu\beta}\nabla^{\nu}h_{\alpha\beta})\xi^{\alpha} \nonumber\\
&\quad + (\nabla_{\alpha}\nabla^{\mu}h-\nabla_{\beta}\nabla_{\alpha}h^{\mu\beta}-\nabla^{\beta}\nabla^{\mu}h_{\alpha\beta}+\Box h_{\alpha}{}^{\mu}+2(R_{\alpha\beta}h^{\mu\beta}+R^{\mu\beta}h_{\alpha\beta})-R_{\alpha}{}^{\mu}h)\nabla^{\alpha}\xi^{\nu} \nonumber\\
&\quad - (2R_{\alpha\beta}\nabla^{\alpha}h^{\beta\mu}-2\nabla_{\alpha}R_{\beta}{}^{\mu}h^{\alpha\beta}-R_{\alpha}{}^{\mu}\nabla^{\alpha}h+\nabla^{\alpha}R_{\alpha}{}^{\mu}h-R_{\alpha\beta}\nabla^{\mu}h^{\alpha\beta}+\nabla^{\mu}R_{\alpha\beta}h^{\alpha\beta})\xi^{\nu} \Big] \nonumber\\
&\quad - [\mu\leftrightarrow\nu],
\end{align}
\begin{align}
k_{\text{Riemann}^2}^{\mu\nu} &=\dfrac{1}{8 \pi G }\Big[\Big(2(R^\mu_{\,\,\alpha\beta\gamma}-R^\mu_{\,\,\beta\alpha\gamma})h^{\nu\gamma}+ R^{\mu\,\,\,\nu}_{\,\, \alpha\,\, \beta}h - R^{\mu\,\,\,\nu}_{\,\,\alpha\,\,\gamma}h_\beta^{\,\,\gamma}-R^{\mu\,\,\,\nu}_{\,\,\beta\,\,\gamma}h_\alpha^{\,\,\gamma} - \nabla^\mu \nabla_\alpha h^\nu_{\,\,\beta}\!\nonumber\\
&+\!\nabla^\mu \nabla_\beta h^\nu_{\,\,\alpha}\Big)\!\nabla^\beta\xi^\alpha +2\Big(R^{\mu\beta}(\nabla_\alpha h^\nu_{\,\,\beta}-\nabla_\beta h^\nu_{\,\, \alpha})+R^{\mu\,\,\,\nu}_{\,\,\beta\,\,\gamma} \nabla^\gamma h_\alpha^{\,\,\beta}+\frac{1}{2}R^{\mu\nu}_{\,\,\,\,\,\alpha\gamma}(\nabla_\beta h^{\beta\gamma}-\nabla^\gamma h)\nonumber\\
&+2(\nabla_\beta R^\mu_{\,\,\alpha}-\nabla^\mu R_{\alpha\beta})h^{\nu\beta}+\nabla^\mu \nabla_\beta\nabla_\alpha h^{\nu\beta}-\nabla^\mu\Box h^\nu_{\,\,\alpha}+\nabla^\mu R^\nu_{\,\,\alpha}h + \nabla^\mu R^\nu_{\,\,\beta} h_\alpha^{\,\,\beta}\nonumber\\
&-\nabla^\mu (R^\nu _{\,\,\beta\alpha\gamma}h^{\beta\gamma})\Big) \xi^\alpha - 2\big(\nabla^\rho R^{\mu}_{\,\,\alpha\rho\beta}h^{\alpha\beta}-R^{\mu}_{\,\,\alpha\rho\beta}\nabla^\rho h^{\alpha\beta}\big) \xi^\nu \Big]-[\mu\leftrightarrow\nu],     
\end{align}
}

\noindent By integrating the exact total variation tensor $k_{\xi}^{\mu\nu}$ over the appropriate field variations, one can reliably derive the conserved charge variation associated with the vector $\xi$.

To proceed, one selects a vector field $\xi^\mu$ together with a metric perturbation $\delta g_{\mu\nu}$ in order to determine the corresponding variation of the conserved charge. The metric perturbation may be taken as a parametric variation \cite{Hajian:2015xlp,Hajian:2014twa}, namely
\begin{equation}
\delta g_{\mu\nu} = \frac{\partial g_{\mu\nu}}{\partial a}\,\delta a.
\end{equation}
For the computation of the angular momenta, we choose $\xi^\mu_i = -\partial_{\varphi^i}$ (equivalently, $\xi^\mu_\phi = -\partial_\phi$ and $\xi^\mu_\psi = -\partial_\psi$).

The same method can also be used to compute the entropy of NHEGs as a conserved charge associated with an infinite family of Killing horizons, each characterized by a bifurcation surface $\mathcal{H}$. For every such surface, there exists a specific Killing vector field $\zeta_{\mathcal{H}}$ \eqref{zetaH} that vanishes on $\mathcal{H}$ and becomes null on the associated Killing horizon. Based on this structure, the entropy $S$ of the solution can be defined as a Noether-Wald charge associated with $\zeta_{\mathcal{H}}$, and its value is independent of the choice of $\mathcal{H}$ due to the underlying SL$(2,\mathbb{R})$ isometry. This leads to two universal thermodynamic laws that characterize the behavior of black holes at zero temperature (reviewed in Ref. \cite{Hajian:2015eha}):
\begin{itemize}
    \item NHEG Entropy Law (also known as Sen's entropy function formula): This law provides a universal relation between the entropy and other thermodynamic variables of the extremal geometry:
    \begin{equation}
        \frac{S}{2\pi} = k^iJ_i - \oint_{\mathcal{H}} \sqrt{-g} \mathcal{L},
    \end{equation}
    where $\mathcal{L}$ is the full Lagrangian density including the GB term. This relation may be viewed as the extremal counterpart of the Smarr relation.
    \item NHEG Entropy Perturbation Law: For field perturbations $\delta g_{\mu\nu}$ that satisfy the linearized equations of motion and preserve the $\{\xi_-, \xi_0\}$ isometry \cite{Hajian:2014twa,Compere:2015mza,Compere:2015bca,Hajian:2017mrf} (e.g., for the parametric variations), the variation of the entropy is governed by:
    \begin{equation}\label{NHEG entropy perturbation}
        \frac{\delta S}{2\pi} = k^i\delta J_i.
    \end{equation}
    This law serves as the extremal counterpart of the first law of black hole thermodynamics, determining the variation $\delta S$ while the Hawking temperature remains fixed at $T_H = 0$.
\end{itemize}

\end{document}